\documentstyle[]{mn}
\input psfig
\def\diff{{\rm d}}
\def\tot{{\rm tot}}
\def\calc{{\rm calc}}
\def\DV{de Vaucouleurs{}}
\def\Se{S\'ersic{}}

\title[Entropy of elliptical galaxies in Coma]%
{The entropy of elliptical galaxies in Coma: a clue for a 
distance indicator\thanks{Based on observations collected at the 
Canada-France-Hawaii telescope, operated by the National Research 
Council of Canada, the Centre National de la Recherche Scientifique of 
France, and the University of Hawaii}}

\author[Gerbal et al.]
 {D. Gerbal,$^{1,2}$ G. B. Lima Neto,$^{3,4}$
  I. M\'arquez,$^{1,5}$ H. Verhagen$^{1,6}$\\
  $^1$ Institut d'Astrophysique de Paris, CNRS, Universit\'e Pierre et 
Marie Curie, 98bis Bd Arago, F-75014 Paris, France\\
  $^2$ DAEC, Observatoire de Paris, Universit\'e Paris VII, CNRS (UA 173), 
F-92195 Meudon Cedex, France\\
  $^3$ Universit\"at Potsdam, c/o Astrophysikalisches Institut Potsdam,
	An der Sternwarte, 16, D-14882 Potsdam, Germany\\
  $^4$ Observatoire de Lyon, Av. Charles Andr\'e, F-69561 St Genis Laval
Cedex, France\\
  $^5$ Instituto de Astrof\'{\i}sica de Andaluc\'{\i}a, 
CSIC, Apdo 3004, 18080 Granada, Spain\\
  $^6$ Sterrenkundig Instituut, `Anton Pannekoek', 
Universiteit van Amsterdam, Nederland}

\pagerange{\pageref{firstpage}--\pageref{lastpage}}
\pubyear{1997}

\begin{document}
\maketitle
\label{firstpage}
\begin{abstract}
We have fitted the surface brightness of a sample of 79 elliptical 
galaxies pertaining to the Coma cluster of galaxies using the \Se\ profile.
This model is defined through three primary parameters:
scale length ($a$), intensity 
($\Sigma_{0}$), and a shape parameter ($\nu$); physical and astrophysical
quantities may be computed from these parameters. We show that correlations
are stronger among primary parameters than the classical astrophysical 
ones. In particular, the galaxies follow a high correlation  
in $\nu$ and $a$ parameters. We show that the $\nu$ and $a$ correlation
satisfies a constant specific entropy condition.
We propose to use this entropy relation as distance indicator for clusters.
\end{abstract}


\section{Introduction}\label{intro}
The photometrical properties of elliptical galaxies have already been 
extensively analysed, with special attention paid to the entire range 
of sizes and luminosities, from giant to dwarf ellipticals.  Profile 
laws are in general defined by various parameters, for example,
scale length ($a$), intensity ($\Sigma_{0}$), and one or more parameters
specifying 
the shape (structure parameters). These parameters can be
categorized as {\em basic} or {\em primary},
in the sense that they come out naturally from a mathematical 
definition.

The \DV\ profile has no structure parameter; for a long time it 
appeared to be in very good agreement with observations, although 
hints of systematic deviations were observed; those deviations appear 
similar for galaxies with the same luminosities (Michard 1985, 
Schombert J.M.  1986).  New observations, especially high resolution 
data from the HST 
(e.g. Ferrarese et al. \shortcite{Ferrarese}), have definitively shown that
the \DV' profile -- rather too inflexible -- does not allow a convenient
description of the actual structure of all elliptical galaxies.

Some attempts have been 
made to obtain more general surface brightness laws.  Among many
possibilities, S\'ersic's recipe \shortcite{Sersic},
\begin{equation}
 \Sigma(r) =  \Sigma_{0} \exp(-u^{\nu})\;,\,  u=r/a\, ,
\label{DVlike} 
\end{equation}
a generalization of the \DV-law, seems to be better suited to describe 
elliptical brightness profiles (e.g. Ciotti \shortcite{Ciotti},
Caon et al. \shortcite{Caon}, Graham et al. \shortcite{Graham},
Courteau et al. \shortcite{Courteau}).

In this letter we show the results of the photometrical analysis of a 
sample of elliptical galaxies in the Coma cluster using a \Se-model 
(or a $\nu$ model).  From the basic parameters [$a$, $\Sigma_{0}$, 
$\nu$], it is possible to build physical and astrophysical quantities 
(effective radius, effective luminosity, total magnitude, energy, 
entropy, etc\ldots).  All these quantities are in fact a combination 
of primary parameters; we therefore propose to use primary parameters 
instead of the classical `astrophysical' quantities in order to obtain 
better defined correlations.

Any correlation between a dimensionless parameter and a 
distance-dependent one has traditionally led to the definition of 
a distance indicator; for instance, the Cepheid luminosity--period 
relation (Luminosity/Period), the Tully-Fisher relation 
(Luminosity/Rotation), and the Faber-Jackson relation 
(Luminosity/Velocity dispersion) are widely used.
Recently, Young \& Currie \shortcite{Young} have 
pointed out a correlation between the (distance-dependent) intrinsic 
luminosity and the (dimensionless) shape parameter of the \Se-law for
dwarf ellipticals.

A theoretical understanding of the physical underlying processes is 
crucial to assure the validity of a correlation between the parameters 
describing an astronomical object.  Here, we show that elliptical 
galaxies are located in the locus defined by constant specific entropy 
$s$, in the primary parameters plane [$\nu$, $a$].
The fact that the elliptical 
galaxies in Comamainly  have the same specific entropy 
should somewhat reflect the intrinsic properties of self--gravitating
systems formed through violent relaxation processes.
We will discuss whether the entropy relation connecting $\nu$ and $a$ may
be used as a distance indicator.

\section{The Sersic--model}\label{model}
\subsection{Scale and structure parameters}

Surface brightness functions generally used to describe elliptical 
galaxies can be written as:
\begin{equation}
\Sigma = \Sigma_{s} \bar{F}(|\vec{\alpha}|;(u))\;,\, u=(r/a)
\label{DF}
\end{equation}
were $\bar{F}$ does not depend on the normalization factor, $\Sigma_{s}$. 
$\Sigma_{s}$ is a 
characteristic intensity and $a$ is a characteristic scale.  Both
are scaling parameters, while the set $\vec{\alpha}$, 
contains the structural parameters which account for the 
specific shape of the surface brightness profile.
The so-called $\beta$--model, $\gamma$--model, or more generally
the ($\alpha$, $\beta$, $\gamma$)--model \cite{Zhao}, are known examples.

From both scaling and structural parameters which are {\em primary} 
in the sense of being directly obtained from the fit, total 
luminosity, effective radius, effective magnitude, etc\ldots 
may be calculated.  Since they result from a combination of the 
primary parameters, they can be considered as {\em secondary 
quantities}.

\subsection{The \Se\ distribution}\label{modelgamma}

To proceed with the fit we have decided to use a general 
brightness profile pertaining to the family described by
Eq. (\ref{DF}), namely the \Se\ model. The fitting function, that we will
refer as the $\nu$--model (due to the analogy with the 
$\beta$-model) is given by Eq.~(\ref{DVlike}).

The relation between $\nu$ and $n$ (which is often used) is 
$n=1/\nu$. For $\nu = 0.25$ (i.e., $n = 4$) the \DV-model is 
recovered. We note that the \DV-model has no structural parameter, 
implying that any galaxy profile can be deduced from a unique galaxy model 
by scaling (homology property).

\subsection{Total and effective quantities}

To calculate the total luminosity and `effective' quantities (i.e.  
quantities related to the `effective' radius $R_{e}$) we have used the 
classical definitions and, after straightforward calculations, we 
find:
\begin{itemize}

 \item for the total luminosity, $L_\tot$:
\begin{equation}
 L_\tot = \Sigma_{0} a^{2} L^{*}(\nu)
 \label{Ltot}
\end{equation}
with:
\begin{equation}
 L^{*}(\nu) = 2 \pi  \frac{1}{\nu} \Gamma(\frac{2}{\nu}) 		
 \label{L*G}
\end{equation}
where $\Gamma(x)$ is the Gamma function. $L^{*}(\nu)$ depends only on the
structure parameter $\nu$. The total magnitude is $m_\tot=-2.5 \log L_\tot +$~cte.
We have approximated the magnitude $m^{*}(\nu)= 
-2.5\log L^{*}(\nu)$ by the following expression:
 \begin{equation}
  m^{*}(\nu)= 2.199  - 1.246\;\nu - {2.720\over\nu} - {0.2195\over\nu^2}
 \label{Mu*gamma}
 \end{equation}
which is accurate to 3\% for $0.1 \leq \nu \leq 1.0$.

 \item for the effective Radius: $R_{e}=a U_{e}$, where 
$U_{e}(\nu)$ depends only on $\nu$; we have 
approximated it by an analytic expression:
\begin{equation}
\ln U_{e}(\nu) =  {1\over\nu}(0.5434 - 1.069 \ln\nu)
	\label{Uegamma}
\end{equation}

 \item for the effective brightness, $\mu_{e}$: 
\begin{equation}
 \mu_{e} = -2.5 \log(\Sigma_{0}) + 1.0857 U_{e}^{\nu}(\nu)
 \label{mueff}
\end{equation}
As it can be seen, $\mu_{e}$ is the sum of two terms; the first one is 
a function of $\Sigma_{0}$ and the second one only depends on 
$\nu$.

 \item For the mean effective brightness, $<\mu>_{e}$:  
\begin{equation}
 <\mu>_{e} = 2.5\;\log2\pi\; -2.5\;\log\Sigma_{0}
 + 5\;\log\; U_{e}(\nu) + m^{*}(\nu)
	\label{mumoyeneff} 
\end{equation}
$<\mu>_{e}$ is the sum of 4 terms, the first being a function of 
$\Sigma_{0}$ while the second and third depend only on $\nu$.

\end{itemize}

Therefore, the photometrical properties of elliptical galaxies can be 
described in terms of a combination of primary parameters. That is to 
say, we may transform the triplet [$\Sigma_{0}, a, \nu$] into the 
triplet [$L_\tot(m_\tot), R_{e}, \nu$]. Correlations 
between the secondary parameters [$L_\tot(m_\tot), R_{e}, \nu$]
should therefore show larger dispersions than those corresponding to 
[$\Sigma_{0}, a, \nu$]. Notice that any strong correlation between either
$\Sigma_0$ or $a$, and $\nu$ should provide a distance indicator.

\section{Analysis of the data and correlations}
\subsection{The data}

The sample of elliptical galaxies has been selected from the catalogue 
by Lobo et al. \shortcite{Lobo}, based on V-band CCD images covering a region 
of $30' \times 21'.6$ centered on the position $\alpha=12^{h} 59^{m} 
42^{s}\kern-.3em.71$, $\delta= +27\degr 58'~14''\kern-.3em.2$,
observed at the 3.6 m 
Canada--France--Hawaii Telescope in 1993.  We have selected 79 galaxies, 
which cover the range [13--17.8] in V-magnitudes and [0.6--20 kpc] in 
effective radius. We assume a distance for Coma of 137 $h^{-1}_{50}$~Mpc
\cite{Colless}. A full description of the sample will be given 
elsewhere \cite{Gerbal}.

\subsection{The fitting method}

We have fitted  the integrated flux (the luminosity growth curve)
as a function of the surface.
We decided to use such a fit since the 
irregularities of the galaxy are smoothed out. While fitting the
light profile is very sensitive to the outer parts, 
fitting the integrated flux gives a better representation of
the global light distribution
(Prugniel \& Simien \shortcite{Prugniel}).
Moreover, in this way we can take into
account the ellipticity of the galaxies.

We therefore write equation (\ref{DVlike}) in an integrated 
form:
\begin{equation}
L_\tot(\varepsilon)=\frac{2b}{\nu}\; \Sigma_{0} \; \hbox{\Large$\gamma$} \! 
\left(\frac{2}{\nu},\left(\frac{\varepsilon}{b}\right)^{\nu/2}\right) 
\; , \label{eq:L(S)}
\end{equation}
where $\gamma(c,x)$ is the incomplete gamma function, $b$ is related to the length
scale $a$ by $b=\pi a^2$, and  $\varepsilon$ is
the surface corresponding to the region for which the flux is higher
than a given level $\Sigma$ (cf. Eq.(\ref{DVlike})). The corresponding mean
radius is then defined as $r= (\pi\varepsilon)^{1/2}$. $L_\tot(\varepsilon)$
is the total flux within a given $\varepsilon$.

The sky level was determined in a similar way as J\o rgensen \& Franx 
\shortcite{Jorgensen}.  In order to limit the influence of the seeing 
(FWHM $\approx 0.9''$) the data points are taken from $3.0''$ 
outwards, out to a surface brightness $\mu_{V}=24.0$ (i.e.  3$\sigma$ 
signal above the background sky noise).  The fitting process has been done 
under Interactive Data Language (IDL) supplying data values with their 
1-$\sigma$ errors.  These are very small ($\sim$ 1\%, 3\%, and 2\% for 
$\nu,\, a,\, \Sigma_0$, respectively) and are correlated with each 
other as 
in the \DV\ or $\beta$-model case.
A more detailed description of the fitting procedure will be given
elsewhere \cite{Gerbal}, together with the resulting values for primary and 
secondary parameters.

\subsection{Correlations between parameters}

\begin{figure}
 \centerline{\psfig{figure=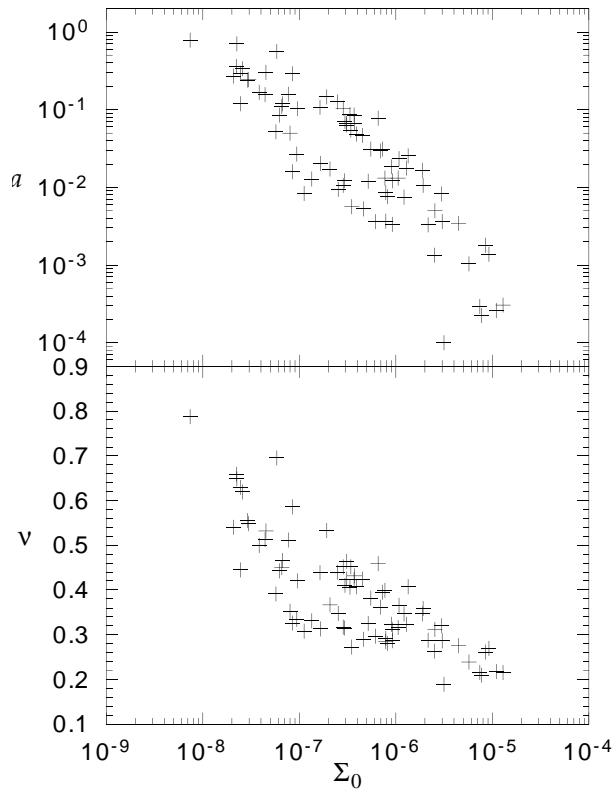,width=8.truecm,clip=true}}
 \caption{Distribution of the primary parameters $a, \Sigma_{0}$ (top), 
 and $\nu, \Sigma_{0}$ (bottom). $\nu$
 is a dimensionless parameter, $a$ is expressed in kpc $h^{-1}_{50}$,
 and $\Sigma_{0}$ in erg~arcsec$^{-2}$~s$^{-1}$ }
	\protect\label{fig1}
\end{figure}

\begin{figure}
	\centerline{\psfig{figure=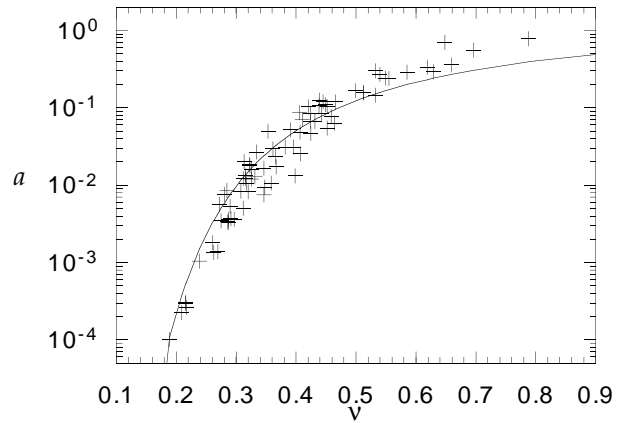,width=8.truecm,clip=true}}
	\caption{Distribution of the primary parameters $a, \nu$. 
	The units are as in Fig. 1. The solid line represents a constant
	specific entropy track (cf. section 4.1).}
	\protect\label{fig2}
\end{figure}

In Fig. \ref{fig1} and Fig. \ref{fig2} we show the relations between
the 3 primary parameters:
$[\nu, \Sigma_{0}]$, $[a, \Sigma_{0}]$, and $[a,\nu]$.

In order to quantify the correlations in Figs. \ref{fig1} and 
\ref{fig2}, we have performed the Spearman and Kendall rank
correlation (non-parametric) tests \cite{Siegel}. The results are
presented in Table \ref{nonparametric}.

\begin{table}
 \caption{Correlation between the parameters of the \Se-law. Both 
 $\rho$ and $\tau$ are defined in the interval $[-1,1]$, the 0 value
 meaning no correlation. Higher absolute values of $z$ indicate a greater
 significance of the correlations}
 \halign{#\hfill&\quad \hfill# &\quad \hfill #\quad
&\quad \hfill# &\quad \hfill #\cr
  \noalign{\medskip\hrule\medskip}
  Parameters&$\rho$-Spearman&$z$-value&$\tau$-Kendall&$z$-value\cr
  \noalign{\smallskip\hrule\medskip}
  $\nu$--$a$        &  0.97 &  8.53 &  0.85 & 11.06\cr
  $\nu$--$\Sigma_0$ & -0.80 & -7.08 & -0.62 & -8.07\cr
  $a$--$\Sigma_0$   & -0.85 & -7.52 & -0.68 & -8.88\cr
  \noalign{\smallskip}
  $\nu$--$R_{e}$    & -0.54 & -4.73 & -0.38 & -4.91\cr
  $\nu$--$m_\tot$  &  0.48 &  4.23 &  0.34 &  4.42\cr
 \noalign{\medskip\hrule\smallskip} 
}	
	\protect\label{nonparametric}
\end{table}

It can be seen that there are correlations between all primary
parameters. However the correlations related to the normalization
factor $\Sigma_{0}$ are less pronounced, while $a$ and $\nu$ are strongly
correlated. The greater $a$, the greater $\nu$, i.e., the greater the 
characteristic length, the steeper the brightness profile.

\begin{figure}
	\centerline{\psfig{figure=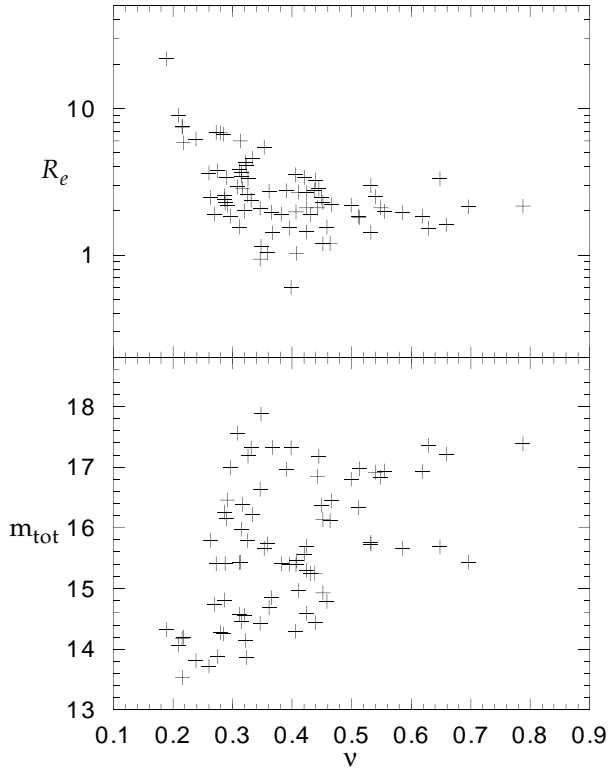,width=8.truecm,clip=true}}
	\caption{Correlations between $\nu$ and two secondary
	 parameters: $ R_{e}$ (in kpc $h^{-1}_{50}$, top) and
    $m_\tot$ (in magnitude, bottom).}
	\protect\label{fig5}
\end{figure}

In Fig.~\ref{fig5} we show the correlations between the secondary 
parameters $[\nu, R_{e}]$ and $[\nu, m_\tot]$.  The secondary 
parameters effective radius, $R_{e}$ and total magnitude, $m_\tot$, 
are clearly much less strongly correlated with $\nu$.

This result could be explained in terms of the noisy combination
of the primary parameters to obtain the second ones and, to a
certain extent, the
role played by $\Sigma_{0}$ (see eq.~\ref{Ltot}),
which is not so well correlated to the two other primary parameters.

\section{Implications of the relation $\bmath{[\lowercase{a}, \nu]}$}

\subsection{A physical interpretation}

As noted by Graham et al. \shortcite{Graham}, the physical origin of
the $[a, \nu]$ 
correlation is not understood up to now.  We will show that this 
relationship has a simple physical interpretation in terms of 
gravo--thermodynamics.  Notice that this kind of relation exists for a 
large family of objects -- from the nuclei of spiral galaxies 
\cite{Courteau} to Brighter Cluster galaxies, and even for the X-ray 
emitting gas in galaxy clusters \cite{Durret}. Therefore, the 
physical explanation should be only related to $a$ and $\nu$ and to a 
small extent to $\Sigma_{0}$.

Gravitational dynamical systems in quasi-equilibrium are in a state of 
quasi-stationary entropy.  We study here the variation of the entropy 
with the parameters $a$ and
$\nu$ of the \Se-law. We use a modified deprojection of the
de Vaucouleurs law derived by Mellier and Mathez (1987) to represent
the 3D-density:
\begin{equation}
\rho(r) = \rho_0 {\left({{r \over a}}\right)}^{-p}
\exp\left({-{\left({{r \over a}}\right)}^{\nu}}\right)\, ,
\label{3D}
\end{equation}
where $\rho_0$ is the normalization parameter.  For any $\nu_{i}$ 
there is one $p_{i}$ so that the $\nu_{i}$--model should be recovered 
by projecting $\rho(r)$ in 2--D.  When $\nu=0.25$ then $p=-0.855$.  In 
order to recover the \Se-law from the projection of Eq.~(\ref{3D}) we 
have derived, following Mellier \& Mathez, the relation:
\begin{equation}
p(\nu) = 0.9976 - 0.5772 \nu + 0.03243 \nu^2\, ,
\label{p(nu)}
\end{equation}
which is accurate to better that 0.1\% in the range $0.1 \le \nu \le 1.0$.

Assuming spherical symmetry, hydrostatic equilibrium, and an isotropic
velocity tensor, it is straightforward to compute the total mass,
$M_\tot$ (assuming a constant mass--to--light ratio), potential
energy, $U$, and pressure profile, $P(r)$. Finally, the entropy
($\cal S$), and the specific entropy ($s$) (following White \&
Narayan \shortcite{White}, assuming an ideal gas), are given by:
\begin{equation}
s = \frac{\cal S}{M_\tot} =
\frac{1}{M_\tot}\int_{V}^{} \ln(\rho^{-5/2}\, P^{3/2})\, \rho\; \diff V\; ,
\label{entropy}
\end{equation}
the integral is over all the volume $V$ of the galaxy.

In Fig.~\ref{fig2} we have plotted the specific entropy as a function 
of $a$ and $\nu$.  As can be expected, the entropy of a 
self-gravitational system of unlimited extent has no maximum.  With 
the present calculation it increases with $\nu$ and with decreasing 
$a$.  Superimposed to the specific entropy curve (isentropic track) 
are the observed values of $a$ and $\nu$ from the fits; one can note 
that they closely follow one another.

%

This result suggests that there should exist some intrinsic relation
between $a$ and $\nu$, the structural parameters of the \Se-law.
However, in order to be able to obtain a maximum entropy in a
dynamical system, one needs to impose more constraints, in addition to
Eq.~(\ref{3D}).
The isentropic track occupied by the observed values of $a$ and $\nu$
may then be 
understood as a maximum entropy class of some more restricted distribution.

\subsection{Towards a distance indicator}

The $\nu$-model, contrary to the \DV-law, does 
not allow to deduce any brightness profile just by scaling; the 
homology property of the \DV-law is lost. Instead, the shape 
property ($\nu$) is related to the scale ($a$) by the relation of 
constant entropy (\ref{entropy}), $s=s_{0}$. Since $a$ depends 
on the distance whereas $\nu$ does not, their correlation can
be used as a distance indicator.

All the galaxies used for the fit belong to the Coma cluster, so they 
are at the same distance (the cluster membership has been 
determined with the redshift, see Biviano et al. 1995), and
thus provide us with a calibration for the distance indicator
based on the $a$-$\nu$ relation.
If the entropy relationship were `universal', i.e., if it were the 
same for different clusters, it could be used as a robust distance 
indicator.  To any observed $\nu_{(k)}$ of any galaxy g$_{k}$ 
in the Universe, would then correspond an $a_{k}$ (in angle units) 
which may be compared to the corresponding value obtained from the 
entropy--relation (Fig.~2), allowing to derive the distance to 
g$_{k}$.

One could wonder if the $a$-$\nu$ relation is weakened by
the correlation between the fitting parameters for a given
galaxy. Since the isentropic track is predicted by theory,
we actually believe that the $a$-$\nu$ relation is robust.

The dispersion of the residue distribution of $\Delta = (a_\calc-a)$ is 
$\simeq 0.1$ leading to a 1--$\sigma$ error of 10\% of the residue.  
The observed dispersion of $\Delta$ could be due to some difficulties 
in the fitting process, or due to physical reasons.  In fact, it seems 
reasonable to imagine that a galaxy which has undergone some energy 
and entropy exchanges with other galaxies (merging, tidal effects, 
etc\ldots) could not verify the entropy relation.

Notice that the $\nu$-profile is a 2-D brightness profile, whereas a 
3-D density profile is needed to calculate the specific entropy.  We 
have used a constant mass-to-light ratio to transform from one to the 
other. Therefore, those galaxies being more luminous than normal ones 
(for instance a galaxy hosting a starburst), will be out of the 
theoretical relationship.

\section{Conclusions}

We have given a physical interpretation for the observed relation 
between two of the {\em primary} parameters of the $\nu$-profile: 
the values of the $a$ and $\nu$ parameters of Coma cluster galaxies
approximately  have a constant specific entropy.  The dispersion around the 
best fit is small, therefore allowing the use of this relation as a 
distance indicator (relative to the Coma cluster).

We must stress that both the observational and theoretical 
evidences found in this work have to be verified.  On the one hand, it 
is necessary to quantify the universality (i.e. in other
clusters) of the ($a$, $\nu$) 
correlation found for the galaxies in the Coma cluster.  On the other 
hand, the validity of the assumptions made to derive the theoretical 
entropy relationship (constant $ M/L$, ideal gas entropy) must be confirmed.

\section{Acknowledgments}
We are very grateful to F. Durret for helpful comments and discussions.
GBLN thanks support from the Alexander von Humboldt Stiftung. IM acknowledges
a post-doctoral fellowship of the Spanish Ministerio de
Educaci\'on y Ciencia.

\label{lastpage}
\end{document}